\documentclass[aps, prd, twocolumn, amsmath, floats,floatfix, superscriptaddress, nofootinbib,
showpacs]{revtex4}

\usepackage{amssymb}
\usepackage{amsmath}
\usepackage{verbatim}
\usepackage{mathrsfs}
\usepackage{amsfonts}
\usepackage{latexsym}
\usepackage{epsfig}
\usepackage{color}
\usepackage{graphicx,subfigure}
\usepackage{units}
\usepackage{slashbox}
\usepackage{envmath}
\usepackage{natbib}

\begin{document}


\definecolor{orange}{rgb}{0.9,0.45,0}

\newcommand{\re}{\mbox{Re}}
\newcommand{\im}{\mbox{Im}}
\newcommand{\tf}[1]{\textcolor{red}{TF: #1}}
\newcommand{\jc}[1]{\textcolor{blue}{JC: #1}}
\newcommand{\nsg}[1]{\textcolor{cyan}{N: #1}}
\newcommand{\ch}[1]{\textcolor{green}{CH: #1}}

\renewcommand{\t}{\times}

\long\def\symbolfootnote[#1]#2{\begingroup%
\def\thefootnote{\fnsymbol{footnote}}\footnote[#1]{#2}\endgroup}


\title{
Dynamical formation of a hairy black hole in a cavity \\  from the decay of unstable solitons} 	

\author{Nicolas Sanchis-Gual}
\affiliation{Departamento de
  Astronom\'{\i}a y Astrof\'{\i}sica, Universitat de Val\`encia,
  Dr. Moliner 50, 46100, Burjassot (Val\`encia), Spain}

\author{Juan Carlos Degollado} 
\affiliation{
Instituto de Ciencias F\'isicas, Universidad Nacional Aut\'onoma de M\'exico,
Apdo. Postal 48-3, 62251, Cuernavaca, Morelos, M\'exico.}

\author{Jos\'e A. Font}
\affiliation{Departamento de
  Astronom\'{\i}a y Astrof\'{\i}sica, Universitat de Val\`encia,
  Dr. Moliner 50, 46100, Burjassot (Val\`encia), Spain}
\affiliation{Observatori Astron\`omic, Universitat de Val\`encia, C/ Catedr\'atico 
  Jos\'e Beltr\'an 2, 46980, Paterna (Val\`encia), Spain}
  
  \author{Carlos Herdeiro}
\affiliation{Departamento de F\'{\i}sica da Universidade de Aveiro and 
Centre for Research and Development in Mathematics and Applications (CIDMA), 
Campus de Santiago, 
3810-183 Aveiro, Portugal}

 \author{Eugen Radu}
\affiliation{Departamento de F\'{\i}sica da Universidade de Aveiro and 
Centre for Research and Development in Mathematics and Applications (CIDMA), 
Campus de Santiago, 
3810-183 Aveiro, Portugal}


\date{November 2016}


\begin{abstract} 
Recent numerical relativity simulations within the 
Einstein--Maxwell--(charged-)Klein-Gordon (EMcKG) system have shown that the 
non-linear evolution of a superradiantly unstable Reissner-Nordstr\"om black 
hole (BH) enclosed in a cavity, leads to the formation of a BH with scalar hair. 
Perturbative evidence for the stability of such hairy BHs has been independently 
established, confirming they are the true endpoints of the superradiant 
instability. The same EMcKG system admits also charged scalar soliton-type 
solutions, which can be either stable or unstable. Using numerical relativity 
techniques, we provide evidence that the time evolution of some of these 
\textit{unstable} solitons leads, again, to the formation of a hairy BH. In some 
other cases, unstable solitons evolve into a (bald) Reissner-Nordstr\"om BH. 
These results establish that the system admits two distinct channels to form 
hairy BHs at the threshold of superradiance: growing hair from an unstable 
(bald) BH, or growing a horizon from an unstable (horizonless) soliton. Some 
parallelism with the case of asymptotically flat boson stars and Kerr BHs with 
scalar hair is drawn.
\end{abstract}




\maketitle

\vspace{0.8cm}

\section{Introduction}

In two recent papers~\cite{Sanchis-Gual:2015lje,Sanchis-Gual:2016tcm}, we have 
investigated the non-linear evolution of the superradiant instability of a 
Reissner-Nordstr\"om (RN) black hole (BH), triggered by a charged scalar field 
confined to a cavity, which surrounds the BH. This system is not only afflicted 
by superradiant instabilities -- unlike asymptotically flat RN 
BHs~\cite{Hod:2013eea, Hod:2013fvl} -- but also allows a sufficiently fast 
development of the 
instability~\cite{Herdeiro:2013pia,Degollado:2013bha,Hod:2015hga,
Hod:2016kpm}, and hence a fully non-linear numerical treatment of the evolution. 
Moreover, within this system, superradiant instabilities occur even within 
spherical symmetry, making their treatment simpler.

Using numerical relativity techniques, we evolved in~\cite{Sanchis-Gual:2015lje,Sanchis-Gual:2016tcm} the 
Einstein--Maxwell--(charged-)Klein-Gordon (EMcKG) system, and showed that the 
endpoint of the instability is a hairy BH of the sort first discussed 
in~\cite{Dolan:2015dha} (see also~\cite{Basu:2016srp}). In the appropriate 
gauge, the scalar hair is a stationary state characterized by a real frequency 
that equals the threshold frequency for the superradiant instability. Thus, 
these BHs \textit{exist at the threshold of superradiance}, making them charged 
analogues of the (rotating) Kerr BHs with 
scalar~\cite{Herdeiro:2014goa,Herdeiro:2015gia} and Proca 
hair~\cite{Herdeiro:2016tmi} discovered in asymptotically flat spacetime (see 
also~\cite{Kleihaus:2015iea,Herdeiro:2015tia,Delgado:2016jxq} for 
generalizations with charge and self-interactions and~\cite{Dias:2011at,Brihaye:2014nba,Herdeiro:2015kha} for analogue solutions in higher dimensions).

\bigskip

Besides the hairy BHs, the same EMcKG system allows for soliton-like solutions, 
which are everywhere regular and 
horizonless~\cite{Ponglertsakul:2016wae,Ponglertsakul:2016anb}. Indeed, the 
hairy BHs may be regarded as a bound state of these solitons with a horizon, in 
the same way that Kerr BHs with scalar (or Proca) hair can be regarded as a 
bound state of rotating boson stars~\cite{Schunck:2003kk} (or rotating Proca 
stars~\cite{Brito:2015pxa}\footnote{Static Proca stars have been recently 
generalized to include charge~\cite{Garcia:2016ldc} or a cosmological 
constant~\cite{Duarte:2016lig}.}) with a horizon (see, 
$e.g.$~\cite{Kastor:1992qy,Herdeiro:2014ima}, for discussions of horizons inside solitons). 
In~\cite{Ponglertsakul:2016wae,Ponglertsakul:2016anb} the linear stability of 
the EMcKG solitons in a cavity was recently addressed and it was shown that some of these 
solutions are unstable against linear spherical perturbations. A natural 
question to ask is what is the end-point of the instability and, in particular, if the 
decay of these unstable solitons could yield a different channel for 
the formation of the aforementioned hairy BHs. 

\bigskip

In this paper we shall address 
these questions, by using similar numerical methods to the ones used
in~\cite{Sanchis-Gual:2015lje,Sanchis-Gual:2016tcm}. As we shall discuss below, 
there are different possible schemes to consider when evolving these EMcKG 
solitons. The reason is that the ``cavity", which is something virtual for the 
stationary soliton solutions, needs to be made concrete, and imposes a concrete 
rule in the dynamical evolution of the system. Different schemes result from a 
different choice of rule, and ultimately, they reflect different dynamical 
problems. We shall illustrate this explicitly, by considering two different 
schemes. For these two possibilities, albeit we observe some quantitative and 
qualitative differences, there is a universal feature: we shall provide 
numerical evidence that indeed a hairy BH at the threshold of superradiance may 
form from the decay of an unstable soliton. But we also observe, for both 
schemes, that some unstable solitons decay into a (bald) RN BH, rather than a 
hairy one.

\bigskip

This paper is organized as follows. We start in section~\ref{section2} by 
briefly describing the stationary solitonic solutions of the EMcKG system in a 
cavity~\cite{Ponglertsakul:2016wae,Ponglertsakul:2016anb}, which we have 
reproduced, and that will provide us with the initial data for our numerical 
evolutions. The framework for these evolutions is briefly reviewed in 
section~\ref{section3}, wherein both schemes are discussed and the 
corresponding numerical results are  presented. Finally, in 
section~\ref{section4}, we provide some further discussion of our results and 
allude to potential issues of our methods. Furthermore, we speculate about the 
implications of our findings to the rotating, asymptotically 
flat case. 

\section{Stationary solitons of the EMcKG system in a cavity}
\label{section2}

\subsection{The framework}
We consider the following action, describing the EMcKG system (here we follow precisely the same conventions as in~\cite{Ponglertsakul:2016wae,Ponglertsakul:2016anb}, which allows a direct comparison with the results therein): $S = \int \sqrt{-g}\mathcal{L}d^4 x$ with
\begin{align}
\mathcal{L}=\frac{R}{2}-\frac{1}{4}F_{\mu \nu  }F^{\mu \nu  } -\frac{1}{2} g^{\mu \nu } D^\ast_{(\mu } \Phi^\ast D^{}_{\nu )} \Phi-\frac{\mu^2}{2}|\Phi|^2 \ ,
\label{eq:action}
\end{align}
where we have used standard notation: $g$ is the metric determinant, $R$ is the Ricci scalar, $\mu$ is the scalar field mass and round brackets denote symmetrization,
$Y_{(\mu \nu )} = \frac{1}{2} \left(Y_{\mu \nu } + Y_{\nu \mu }\right)$ for a tensor field $Y_{\mu \nu }$.
We use a mostly positive space-time signature, and units in which $8\pi G = c=1$.
The scalar field $\Phi $ is complex, and $\Phi ^{\ast }$ is the complex conjugate of $\Phi $.
The electromagnetic field strength is denoted $F_{\mu \nu } = \nabla_\mu A_{\nu } - \nabla_\nu A_{\mu}$,
where $A_{\mu }$ is the electromagnetic potential.
In (\ref{eq:action}), we have introduced the gauge covariant derivative, $
D_{\mu } = \nabla_{\mu } - i q A_{\mu }$, where $\nabla _{\mu }$ is the covariant derivative and $q$ is the scalar field charge. These conventions are slightly different from the ones in our previous works~\cite{Sanchis-Gual:2015lje,Sanchis-Gual:2016tcm}. We shall further remark on this in section~\ref{section3}. Observe that the EMcKG system is invariant under the gauge transformation:
\begin{equation}
A_\mu \rightarrow \tilde{A}_\mu=A_\mu+\partial_\mu \chi(x) \, , \qquad \Phi\rightarrow \tilde{\Phi}=e^{iq\chi(x)}\Phi \ ,
\label{gt}
\end{equation}
for a real function $\chi(x)$.

\subsection{The solitons}

To obtain the stationary solitonic solutions~\cite{Ponglertsakul:2016wae,Ponglertsakul:2016anb},
we shall consider an ansatz describing spherical, time-independent configurations.
For an isotropic coordinate system, the metric ansatz reads
\begin{equation}
ds^2=-f_0(r)dt^2+f_1(r)[dr^2+r^2(d\theta^2+\sin^2\theta d\varphi^2)] \ ,
\label{ansatzg}
\end{equation}
while the matter fields are
\begin{equation}
A=A_0(r) dt\ ,~~\Phi=\phi(r)\ .
\label{ansatzaphi}
\end{equation}
In the following we shall be working with ansatz~\eqref{ansatzg}-\eqref{ansatzaphi}. But we observe that the gauge transformation~\eqref{gt}, with $\chi=-\omega t/ q$ yields the fields:
\begin{equation}
\tilde{A}=\left(A_0(r)-\frac{\omega}{q}\right) dt\ ,~~\tilde{\Phi}=e^{-i\omega t}\phi(r)\ .
\end{equation}
In this gauge the scalar field oscillates with frequency $\omega$. We recall that the critical frequency for superradiance, $\omega_c$, around a RN BH, is set  by the horizon electric potential as $\omega_c=q\phi(r_H)\equiv -q\tilde{A}_0(r_H)$, where $r_H$ is the radial coordinate of the horizon. This is the condition obeyed by a BH at the threshold of superradiance, which in terms of the `old' gauge~\eqref{ansatzaphi} reads $A_0(r_H)=0$.

With our metric ansatz the EMcKG equations reduce to a set of four second order equations plus
a constraint (which is one of the Einstein eqs.).
For solitons, these equations are solved starting with the
following small-$r$ expansion
\begin{eqnarray}
\nonumber
&&
f_0(r)=f_{00}+\frac{1}{6}(2 a_0^2 q^2 -f_{00} \mu^2)f_{10} \phi_0^2 r^2+\mathcal{O}(r^4)\ ,
\\
\nonumber
&&
f_1(r)=f_{10}-\frac{f_{10}^2}{12f_{00}}(a_0^2 q^2 +f_{00}\mu^2)\phi_0^2 r^2+\mathcal{O}(r^4)\ , \ \ 
\\
\label{expansion}
&&
A_0(r)=a_0+\frac{1}{6}a_0  f_{10 }q^2 \phi_0^2 r^2+\mathcal{O}(r^4) \ ,
\\
\nonumber
&&
\phi(r)=\phi_0-\frac{ f_{10}}{6f_{00}}(a_0^2 q^2-f_{00}\mu^2)\phi_0  r^2+\mathcal{O}(r^4)\ ,
\end{eqnarray}
with $f_{00}, f_{10}, a_0, \phi_0$ four arbitrary constants and
$f_{00}, f_{10}$ strictly positive such that the metric has the
correct signature.
In the numerical integration one takes $f_{00}=f_{10}=1$.
Then a continuum of solutions is found by varying the remaining parameters $ a_0,\phi_0$
together with $q,\mu$. 

\begin{figure}[t]
\begin{minipage}{1\linewidth}
  \vspace{-0cm}\includegraphics[width=1.03\textwidth]{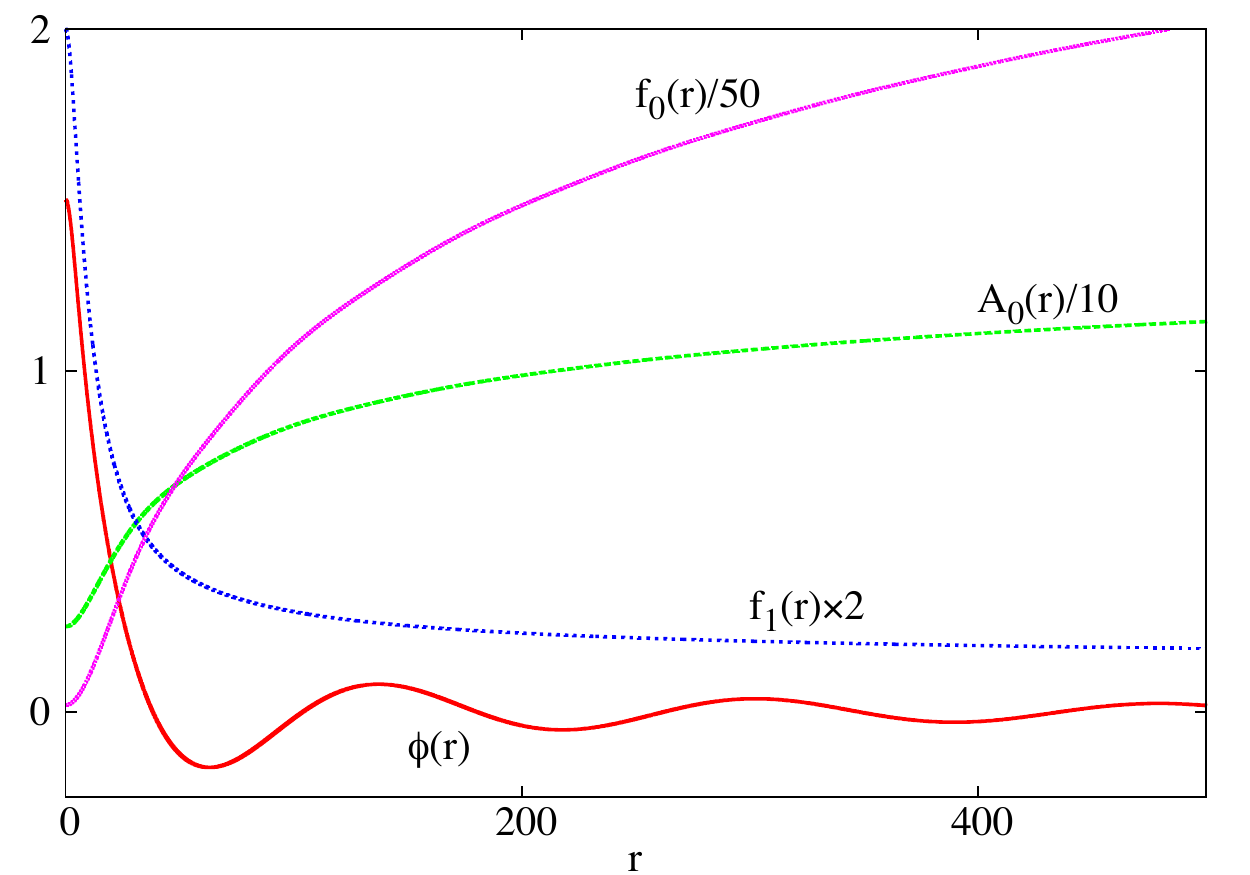}
  \hspace{-0.5cm}\includegraphics[width=1.03\textwidth]{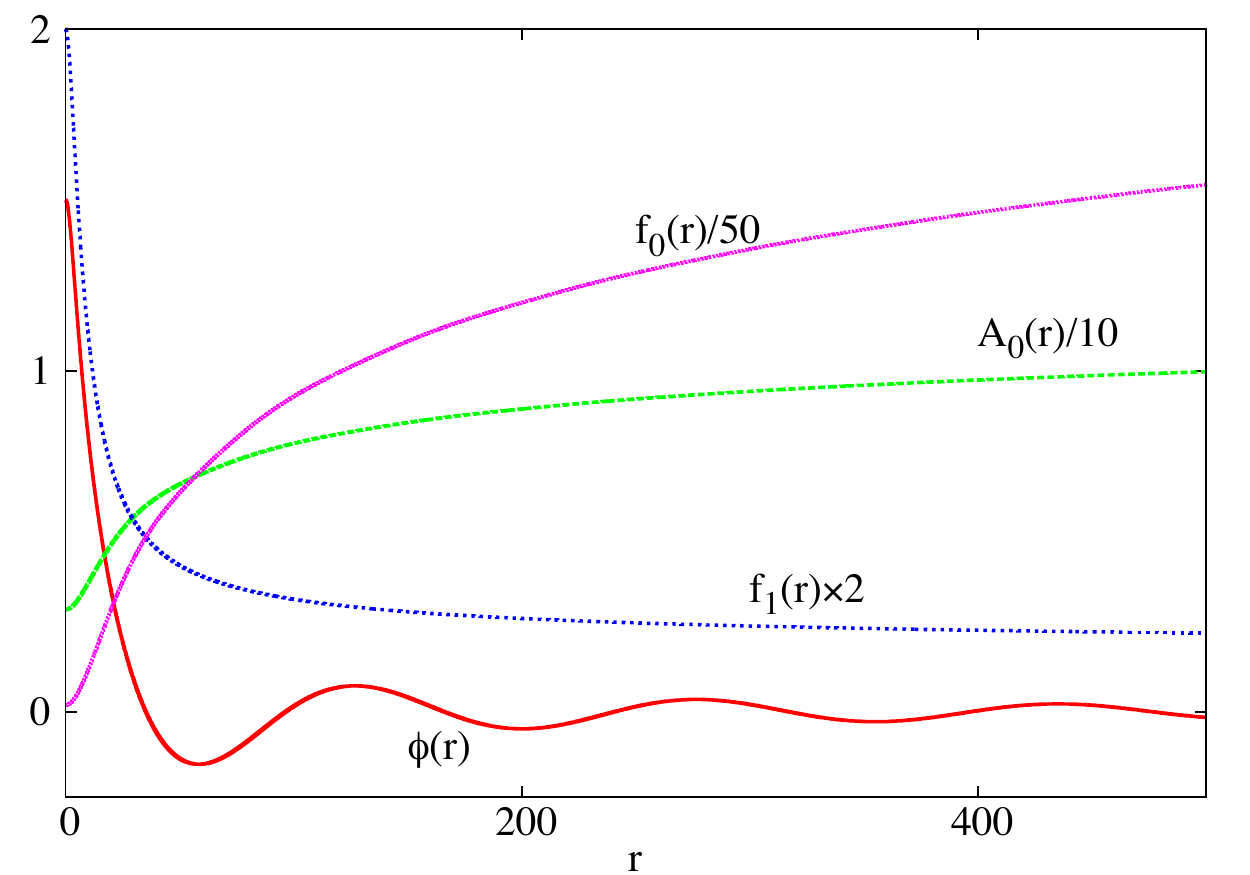}
\caption{Metric, scalar field and gauge field functions for model 1 
(top panel) and model 5 (bottom panel). Some functions are rescaled for better visualization. Observe the oscillations of the scalar 
field. The mirror position is at the first zero of the scalar field (see 
Table~\ref{table1}).}
\label{fig:sol}
\end{minipage}
\end{figure}

Following \cite{Ponglertsakul:2016wae} we choose $q=0.1$. We shall also focus on the massless scalar field case $\mu=0$, but some remarks about the massive case will be made in section~\ref{section4}.
The  field equations are integrated numerically
to find solitonic solutions by using a standard Runge-Kutta method. 
We start the numerical
integration at $r=\epsilon$, where $\epsilon$ is typically $10^{-8}$, using
the expansions (\ref{expansion}) as initial conditions.
One finds that the scalar
field $\phi$ oscillates about zero (see Fig.~\ref{fig:sol}); 
the mirror defining the cavity can be placed at
any zero of the scalar field, but here we shall always consider the mirror to be located at the first zero of the scalar field, whose radial position is labelled $r=r_m$. Thus, in this model, for the stationary solutions, the mirror is ``virtual" rather than corresponding to a concrete physical object that impacts on the EMcKG equations.

In section~\ref{section3} we shall evolve eleven initial data sets, namely one which, according to~\cite{Ponglertsakul:2016wae}, is associated with a stable solution (Model 0) and ten which are associated with unstable solitons (Models 1-10), all with $q=0.1$ and $\mu=0$.\footnote{That these are stable or unstable can be observed by examining Fig. 11 in~\cite{Ponglertsakul:2016wae}.} These models are summarized in Table~\ref{table1}, where we also anticipate the outcome of their numerical evolution using our two schemes. To exemplify the behaviour of the various functions $\{f_0(r), f_1(r), \phi(r),A_0(r)\}$ for typical solitons, we display in Fig.~\ref{fig:sol} their radial dependence for models 1 and 5 in Table~\ref{table1}. According to the results of~\cite{Ponglertsakul:2016wae} these are both unstable solitons. Indeed, as we discuss below, we see them evolving in both schemes. In particular, for scheme I, one of these will be shown to decay into a hairy BH (model 1) and the other to decay into a bald BH (model 5).

\begin{table*}
\begin{tabular}{| c | c | c | c || c | c| c|| c| c| c|| c| c| c| c| }
\hline
Model    &  $a_0$ & $\phi_0$ & $r_m$ & Scheme I & $\omega_c$ & $\omega_\Phi$ &  Scheme II a & $\omega_c$ & $\omega_\Phi$&  Scheme II b & $\omega_c$ & $\omega_\Phi$\\
   \hline
   \hline
   0 & 1.5  & 1.5 & 45.2 &  HBH  & 0.044 & 0.050 & Stable & & & Stable & &  \\
   \hline
1  & 2.5  & 1.5 & 38.2 & HBH & 0.057 & 0.056 & (bald) RN & --- & --- &(bald) RN & --- & ---\\
2  & 2.6  & 1.5 & 37.6 & HBH & 0.057 & 0.055 & (bald) RN & --- & --- &(bald) RN & --- & ---\\
3  & 2.7  & 1.5 & 37.0 & HBH & 0.058 & 0.055 & (bald) RN & --- & --- &(bald) RN & --- & ---\\
4  & 2.8  & 1.5 & 36.4 & HBH & 0.056 & 0.054 & (bald) RN & --- & --- &(bald) RN & --- & ---\\
5  & 3.0  & 1.5 & 35.2 & (bald) RN & --- & --- & (bald) RN & --- & --- &(bald) RN & --- & ---\\
\hline
6  & 3.0  & 1.8 & 56.8 & HBH & 0.040 & 0.039 & HBH & 0.004 & 0.004 & HBH  & 0.004 & 0.004\\
7  & 3.2  & 1.8 & 55.8 & HBH & 0.042 & 0.042 & HBH & 0.005 & 0.004 & HBH & 0.004 & 0.004 \\
8  & 3.4  & 1.8 & 54.8 & HBH & 0.043 & 0.042 & HBH & 0.006 & 0.005 & HBH  & 0.004 & 0.004\\
9  & 4.0  & 1.8 & 52.0 & HBH & 0.048 & 0.046 & (bald) RN & --- & --- & (bald) RN  & --- & ---\\
10 & 4.5  & 1.8 & 49.8 & (bald) RN & --- & --- & (bald) RN & --- & --- & (bald) RN & --- & --- \\
\hline
\end{tabular}
\caption{The eleven soliton models we use as initial data, all with $q=0.1$ and $\mu=0$. From left to right the columns report: the model number, the parameters $a_0$, $\phi_0$ and $r_m$ characterizing the corresponding solitons (columns 2 to 4), and the outcome of the evolutions for the two schemes (following columns), in particular if a hairy BH (HBH) forms or if a bald RN BH results from the evolution. In the former case, the critical frequency of the BH, $\omega_c$, and the corresponding frecuency of oscillation of the scalar field, $\omega_\Phi$, are also indicated.} 
\label{table1}
\end{table*}

\section{Numerical evolutions}
\label{section3}

\subsection{The framework}

The numerical evolutions of our initial data are performed in the EMcKG system~\eqref{eq:action} but, following our previous work~\cite{Sanchis-Gual:2015lje,Sanchis-Gual:2016tcm}, we use slightly different conventions, corresponding to the following rescaling of the fields shown in~\eqref{eq:action}:
\begin{equation}
A_\mu\longrightarrow \sqrt{2} A_\mu \ , \qquad \Phi\longrightarrow \sqrt{8\pi}\Phi \ , \qquad q\longrightarrow \frac{q}{\sqrt{2}} \ .
\label{rescaling}
\end{equation}
The metric and mass parameter are unchanged. Rescalings~\eqref{rescaling} are imposed to all the data in Table~\ref{table1}, which the table reports before the rescaling. This change also ensures that we use the same values for $a_{0}$, $\phi_{0}$ and $q$ than in \cite{Ponglertsakul:2016wae}.

The time update of the different systems of evolution equations we have to solve in our code 
(Einstein, Klein-Gordon, and Maxwell) is done using the same type of numerical techniques we have extensively 
used in previous work (see, in particular,~\cite{Sanchis-Gual:2016tcm,Montero:2012yr,Sanchis-Gual:2015bh,Sanchis-Gual:2015sms}). We refer the interested reader to those references for full details. Here, we simply mention that the evolution equations are integrated using 
the second-order PIRK method developed by \cite{Isabel:2012arx,Casas:2014}.  This 
method allows to handle the singular terms that appear in the evolution equations due to our choice 
of curvilinear coordinates. The derivatives in the spacetime evolution are computed using a 
fourth-order centered finite-difference approximation on a logarithmic radial grid.  
We also use fourth-order Kreiss-Oliger dissipation to avoid high-frequency noise appearing near the 
outer numerical boundary. For the simulations presented in this work 
we evolve the electric field explicitly and the electric potentials implicitly.

We perform the evolution of the unstable and the stable solitons by rescaling 
the lapse function of the 3+1 formalism \cite{Alcubierre08a} in the 
following way:
\begin{equation}
\tilde\alpha=\alpha/\alpha(r_{\text{max}}).
\end{equation}
where $\alpha=\sqrt{f_0}$.
The lapse is rescaled to facilitate our numerical simulations, as it ensures that
$\tilde\alpha\rightarrow1$ at the outer boundary of our grid, which corresponds to Minkowski spacetime.

\subsection{The two schemes}

We are interested in the dynamics of the solitons inside the cavity bounded by 
the mirror at $r=r_m$. Whereas the presence of the mirror in the stationary 
soliton solution is a matter of perspective, its presence in a dynamical 
evolution must 
be enforced by some dynamical rule. In other words, since a generic evolution of 
the initial data provided by the solitons described in the previous section, 
will not preserve the node at $r=r_m$, one has to place a mirror at the location 
of the node, enclosing a cavity. Here we shall consider two different dynamical 
rules.

In \textit{Scheme I} we shall use the initial data provided by the stationary 
soliton solutions \textit{only} inside the cavity. That is, scheme I is based on 
a \textit{truncated soliton}. Outside the cavity we change the initial data by 
setting the scalar field to zero all along the computational domain, but leaving
the metric fields unchanged. Whereas 
this is the simplest thing to do, within the ``scalar field enclosed in a 
cavity" perspective, it has the obvious shortcoming that the initial 
data outside the cavity does not satisfy the field equations. In other words, 
the initial data is constraint-violating outside the cavity. 
As we show below, however, the simulation converges  to a 
solution of the field equations within acceptable errors for this system, the 
endpoint being either a hairy BH or a bald BH, depending on the starting soliton. 
We remark that this dynamical perspective departs from that encoded in the 
perturbation theory considered in~\cite{Ponglertsakul:2016wae}, where the 
soliton extends to spatial infinity and the perturbations are required to 
preserve the node at the mirror location. As a consequence of this difference, 
model 0 in Table~\ref{table1}, which according to~\cite{Ponglertsakul:2016wae} 
is a stable soliton, turns out to evolve in scheme I. 

 In \textit{Scheme II}, on the other hand, we shall use the initial data provided by the stationary soliton solutions \textit{both} inside and outside the cavity. We therefore use the \textit{full soliton}. The presence of the mirror is enforced by not evolving the scalar field value at the mirror, which is initially put to zero. This ensures the presence of the node during the whole evolution and (within numerical error) the absence of scalar energy or charge flux between the inside and outside of the cavity. This enforcement leads to more localized  constraint violations, which occur only at the mirror location, and to a closer correspondence with the perturbation problem discussed in~\cite{Ponglertsakul:2016wae}. The results of the evolutions, however, are qualitatively similar to those of scheme I, $i.e.$ we see two possible outcomes, \textit{but} the ``same" soliton in both schemes may have a different fate. In scheme II we perturb the soliton inside the cavity by multiplying $\Phi$ by a factor slightly larger than one in order to trigger the evolution; then we consider two different perturbation factors leading to two sub-schemes: scheme IIa, with a factor of 1.05, and scheme IIb, with a factor of 1.01.
 
Our simulations employ a logarithmic radial grid with a maximum resolution of
$\Delta r=0.025M$. The outer boundary of the computational domain is placed at 
$r_{\rm{max}}=4.5\times10^3$, far enough as to not affect the dynamics in the inner region during the entire extent of the
simulations. The (Courant) time step is given by $\Delta t=0.1\Delta r$ which ensures long-term stable simulations.

\subsection{Numerical results - scheme I}

\begin{figure}[h!]
\begin{minipage}{1\linewidth}
  \vspace{-0cm}\includegraphics[width=1.03\textwidth]{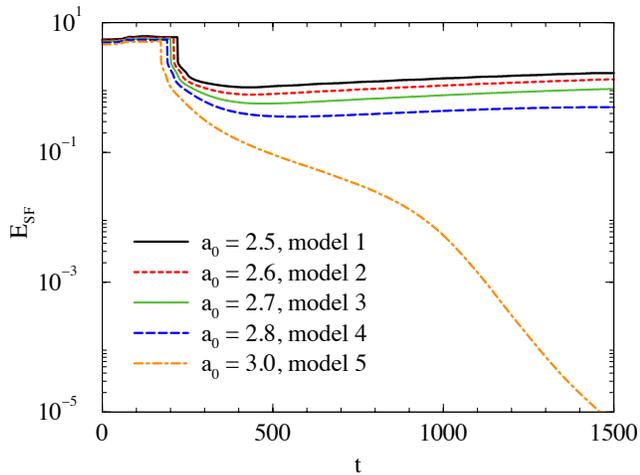}
\vspace{-0cm}\includegraphics[width=1.03\textwidth]{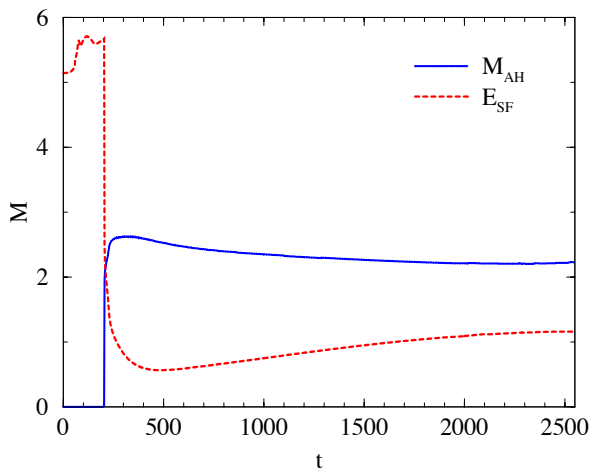}
\caption{Time evolutions in scheme I: the top panel shows the scalar field 
energy  for models 1-5, while the bottom panel shows the AH mass and scalar 
field energy for model 3.}
\label{fig:graf1}
\end{minipage}
\end{figure}

Let us start by describing the numerical results under scheme I. In Fig.~\ref{fig:graf1} (top panel) we show the result for the evolution of the energy in the scalar field computed as in~\cite{Sanchis-Gual:2015lje,Sanchis-Gual:2016tcm}, for models 1-5. We observe that after an initial period of no remarkable change, all models exhibit a sharp decrease of the energy in the scalar field at $t\gtrsim 150$. For models 1-4 what follows is a new, non-zero, equilibrium value of the scalar field energy. However, for model 5 the energy in the scalar field keeps decreasing and tends to vanish asymptotically.
\begin{figure}[h!]
\begin{minipage}{1\linewidth}
  \vspace{-0cm}\includegraphics[width=1.03\textwidth]{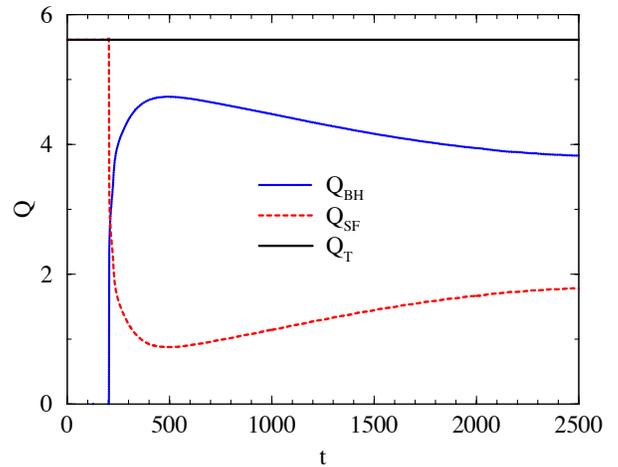}
    \vspace{-0cm}\includegraphics[width=1.03\textwidth]{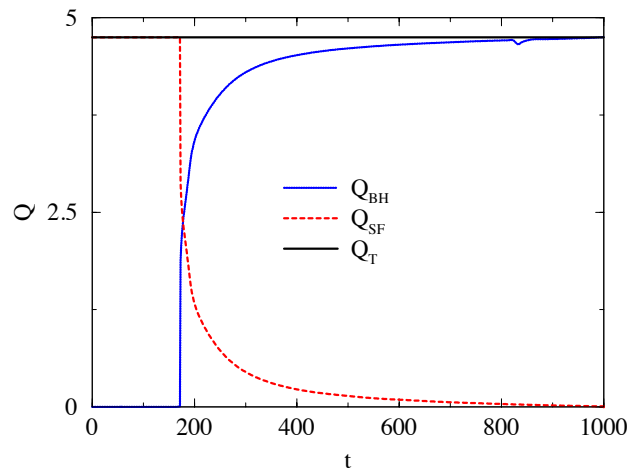}
\caption{Time evolution of the charge in the scalar field, $Q_{\rm SF}$, in the BH (computed at the AH), $Q_{\rm BH}$, and of the total charge, $Q_{\rm T}$, for models 3 (top panel) and 5 (bottom panel), in scheme I.}
\label{fig:graf2}
\end{minipage}
\end{figure}

In all models 1-5 we observe the formation of an apparent horizon (AH) at a time which roughly coincides with the sharp decay of the scalar field energy visible in Fig.~\ref{fig:graf1}. This is illustrated in the bottom panel of the figure for model 3. One can then compute the charge inside the AH and the charge in the scalar field using the expressions described in~\cite{Sanchis-Gual:2015lje,Sanchis-Gual:2016tcm}. The result is exhibited in  Fig.~\ref{fig:graf2} for models 3 and 5. A clear charge transfer from the scalar field (soliton) to the BH is observed, with the total charge (black solid line) remaining constant within an excellent approximation. Whereas in model 3 the charge in the scalar field remains non-zero signaling the formation of a hairy BH, in model 5 the scalar field charge vanishes, signaling the formation of a (bald) RN BH.

\begin{figure}
\begin{minipage}{1\linewidth}
  \vspace{-0cm}\includegraphics[width=1.03\textwidth]{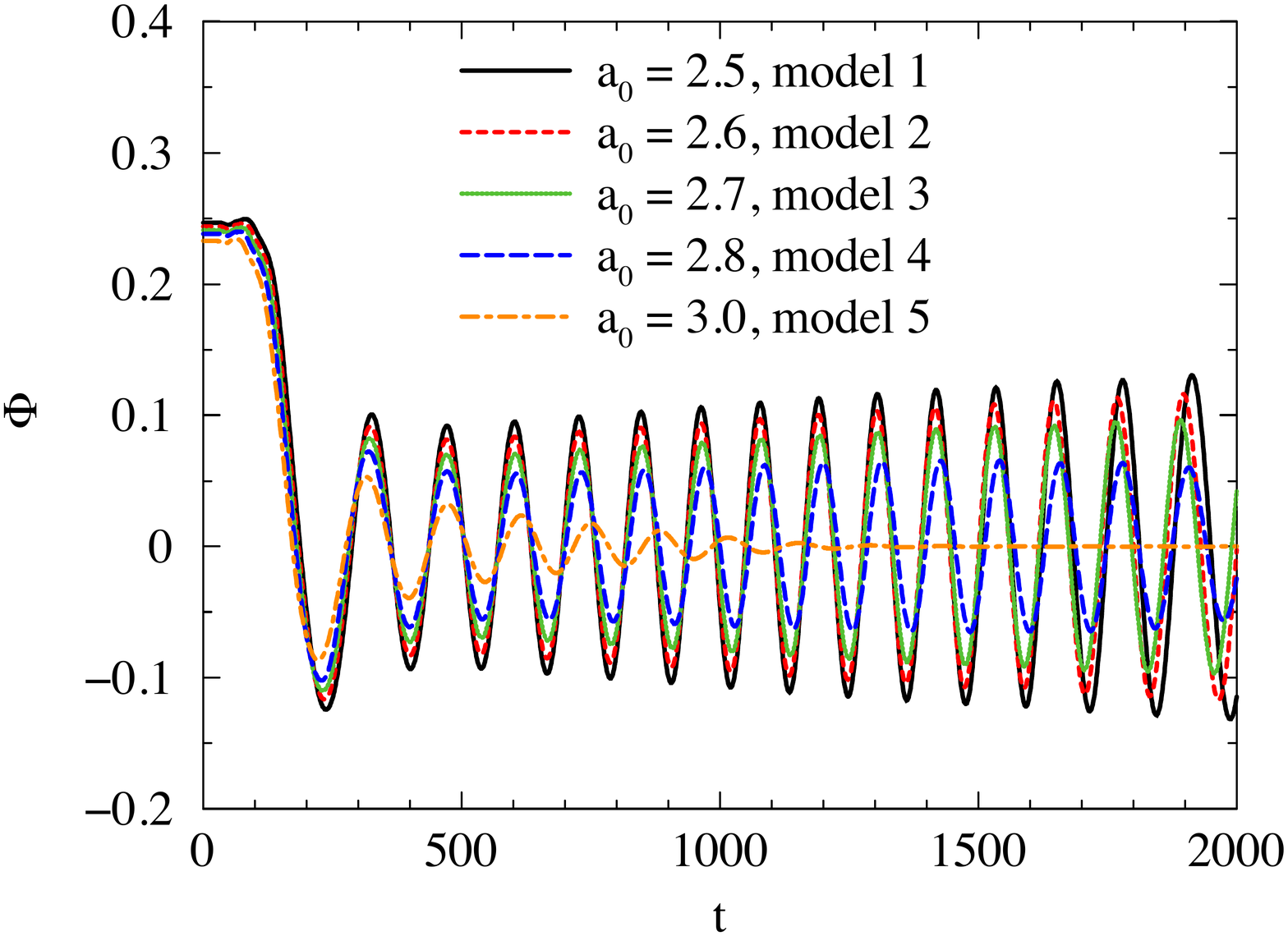}
  \vspace{-0cm}\includegraphics[width=1.03\textwidth]{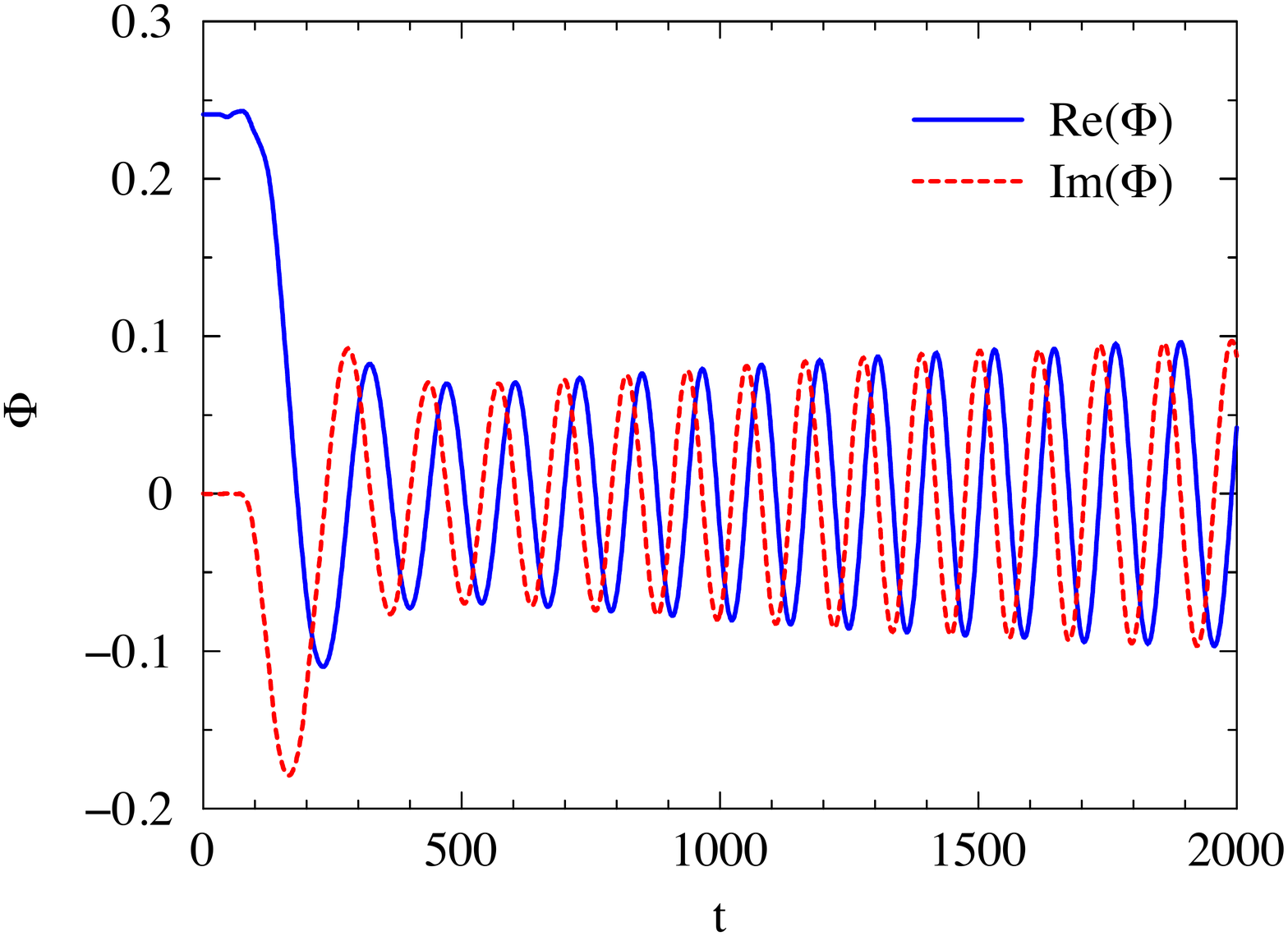}
\caption{Time evolution of the real part of the scalar field amplitude for models 1-5 (top panel) and of the real (blue solid line) and imaginary (red dashed line) parts of the scalar field amplitude for model 3 (bottom panel). Data are extracted at $r=5$. Notice the decay of the field amplitude in model 5.}
\label{fig:graf3}
\end{minipage}
\end{figure}

To examine in more detail the type of hairy BH that forms in the decay of the soliton we show in the top panel of~Fig.~\ref{fig:graf3} the real part of the scalar field for models 1-5, as a function of time, extracted at a specific radial position $r=5$. Correspondingly, the bottom panel of~Fig.~\ref{fig:graf3} displays the real and imaginary parts of the scalar field for model 3  as an illustrative example. It can be seen that for all these models the real part is a sinusoidal function and that the real and imaginary parts have a phase difference of $\pi/2$; therefore, they oscillate with an opposite phase. The critical frequency for the BH that forms and the frequency of oscillation of the scalar field, computed as in~\cite{Sanchis-Gual:2015lje,Sanchis-Gual:2016tcm} are shown in Table~\ref{table1}. They match within the numerical error, showing that these BHs are hairy BHs at the threshold of superradiance.  

\begin{figure}
\begin{minipage}{1\linewidth}
  \vspace{-0cm}\includegraphics[width=1.03\textwidth]{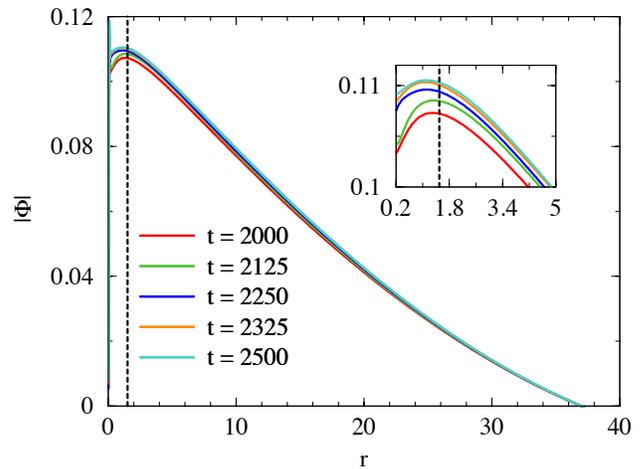}
\caption{Radial profile of the scalar field magnitude for model 3 at different times of the evolution with scheme I. The vertical dashed line marks the location of the AH.
}
\label{fig:graf4}
\end{minipage}
\end{figure}

Finally, we display in Fig.~\ref{fig:graf4}  the radial profile of the scalar field magnitude for model 3 and at different times of the evolution. It is worth noticing that the maximum value attained by the scalar field (best visible in the inset) is inside the BH horizon (signaled by the dashed vertical line). Therefore, the scalar field decays monotonically from the horizon to the mirror, as one expects for stable hairy BHs~\cite{Dolan:2015dha,Sanchis-Gual:2015lje}. 

\begin{figure}
\begin{minipage}{1\linewidth}\hspace{1.5cm}\includegraphics[width=1.03\textwidth, height=0.3\textheight]{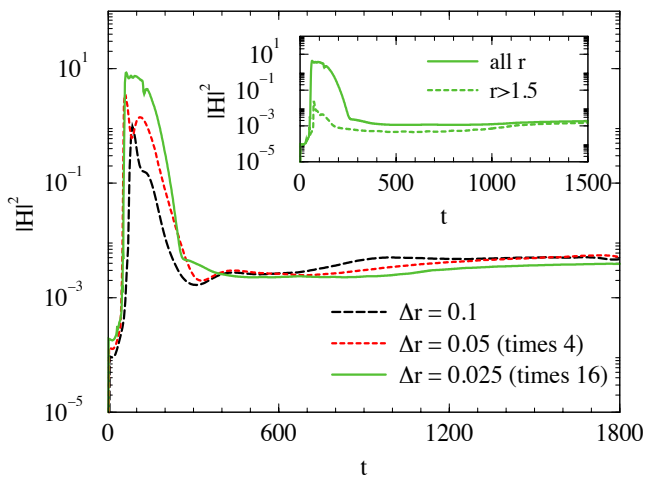}
\caption{Time evolution of the L2 norm of the Hamiltonian constraint for model 3 using scheme I. 
The inset shows the evolution when considering the entire radial domain (solid line) and when removing the radial
zones that eventually lie within the AH (dashed line).
Second-order convergence is  approximately achieved in the evolution for $t\gtrsim 400$.
}
\label{fig:graf5}
\end{minipage}
\end{figure}

The results we have just discussed establish that a hairy BH can form as the outcome of numerical evolutions of unstable solitons. There are, however, two issues that must be discussed in the current scheme. The first issue is related to the constraint violations of the simulations. These arise due to the choice of initial data in scheme I, in which we set the scalar field to zero outside the mirror, but also due to the coordinate singularity at $r=0$. Figure \ref{fig:graf5} plots the time evolution of the L2 norm of the Hamiltonian constraint for model 3 for three different radial grid resolutions. Whereas relatively large constraint violations are observed at some early phase of the evolution, namely during the formation of the AH and shortly after, the values of the violations decrease significantly for the remaining part of the evolution. We note that the values attained in the final state are comparable to those observed in our previous works~\cite{Sanchis-Gual:2015lje,Sanchis-Gual:2016tcm}. 

\begin{figure}[t]
\begin{minipage}{1\linewidth}
  \vspace{-0cm}\includegraphics[width=1.03\textwidth]{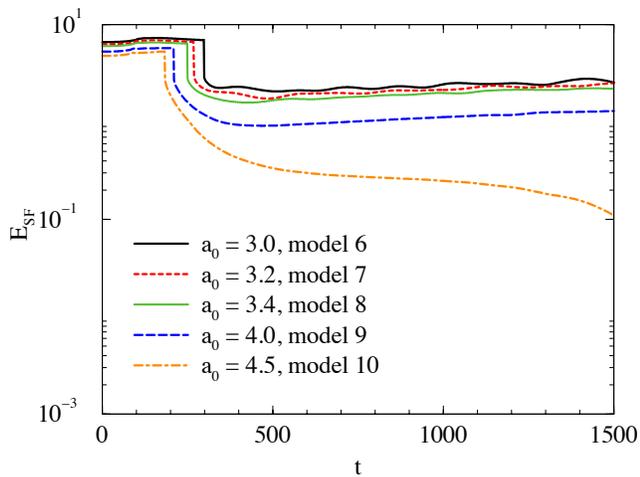}
\vspace{-0cm}\includegraphics[width=1.03\textwidth]{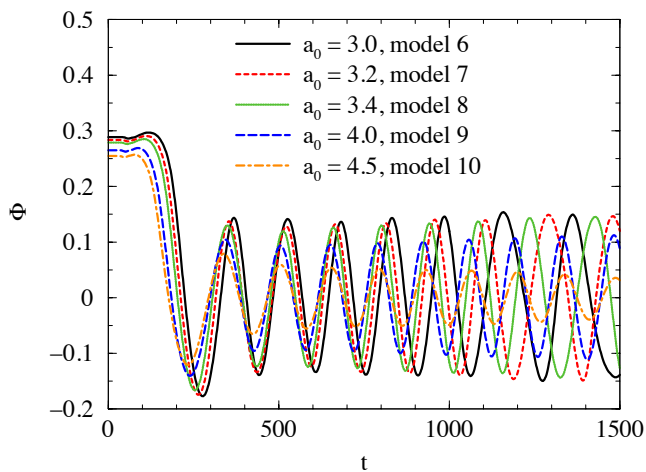}
\caption{Time evolution of the scalar field energy (top panel) and of the real part of the scalar field amplitude (bottom panel)  for models 6-10 in scheme I. The extraction radius is at $r=5$.}
\label{fig:graf6}
\end{minipage}
\end{figure}

\begin{figure}[t]
\begin{minipage}{1\linewidth}
  \vspace{-0cm}\includegraphics[width=1.03\textwidth]{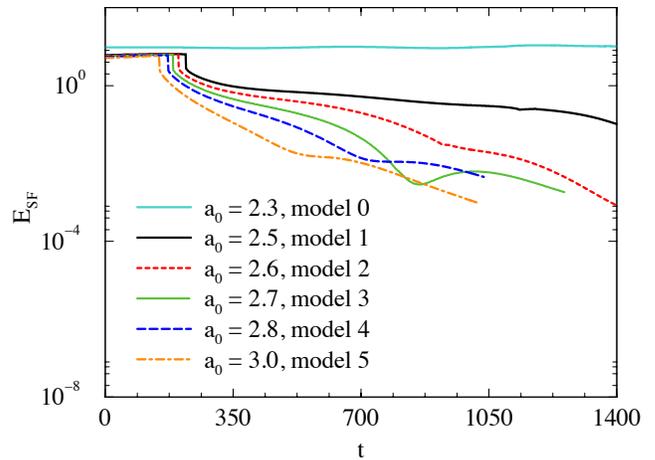}
\vspace{-0cm}\includegraphics[width=1.03\textwidth]{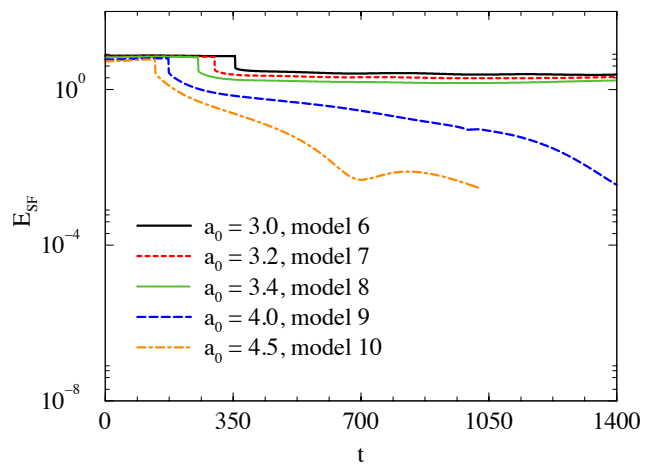}
\caption{Time evolutions of the scalar field energy for models 0-5 (top panel) and models 6-10 (bottom panel), using scheme IIa.}
\label{fig:graf7}
\end{minipage}
\end{figure}

The second issue is related to the stable solution - model 0 in Table~\ref{table1}.  We have checked that   the time evolution of this model indicates it is not stable; rather it evolves. What concerns us here is not so much what is the endpoint of this evolution, but the fact that the soliton in our dynamical scheme I is not stable, whereas it is stable in the perturbation scheme considered in~\cite{Ponglertsakul:2016wae,Ponglertsakul:2016anb}. This discrepancy underlines the fact that the dynamical perturbations considered therein are not faithfully represented by our scheme I, and leads us to consider scheme II below. 

To argue the generality of the behaviour observed for models 1-5 under scheme I,  we display in figure \ref{fig:graf6} the time evolution of the scalar field energy (top panel) and the oscillations of the real part of the scalar field (bottom panel) for models 6-10.  We observe the same trend to that found in models 1-5, namely, for the same value of $\phi_0$ sufficiently small values of $a_0$  lead to a hairy BH, but for larger values  the final state changes and a bald RN BH is obtained instead. Model 10 is the only one whose energy decreases during the rest of the evolution after the BH is formed. The total amount of scalar field energy remaining in the cavity is larger than for the previous five models.

\subsection{Numerical results - scheme II}

We now turn our attention to scheme II.  In Fig.~\ref{fig:graf7} we plot the evolution of the scalar field energy, for models 0-10 within scheme IIa (the first six models in the top panel and the last five in the bottom one). The first observation is that model 0 is now stable, in accordance to~\cite{Ponglertsakul:2016wae}, and the energy remains almost constant, taking into account that the soliton is slightly perturbed. For the unstable models 1-5, a BH forms at $t\gtrsim 200$ absorbing part of the scalar field energy. For all these models, the energy in the scalar field keeps decreasing with time, tending to vanish rather than approaching an equilibrium state. A different result is found for models 6-8, displayed in the bottom panel of Fig.~\ref{fig:graf7}. For these models the scalar field energy is almost constant after the formation of the BH. As observed in scheme I, however, increasing $a_{0}$ for fixed $\phi_0$ beyond a certain value,  the scalar field energy keeps decreasing with time, indicating that a bald RN BH, rather than a hairy BH, forms (cf. models 9 and 10 in the bottom panel of Fig.~\ref{fig:graf7}).

\begin{figure}[t]
\begin{minipage}{1\linewidth}
    \vspace{-0cm}\includegraphics[width=1.03\textwidth]{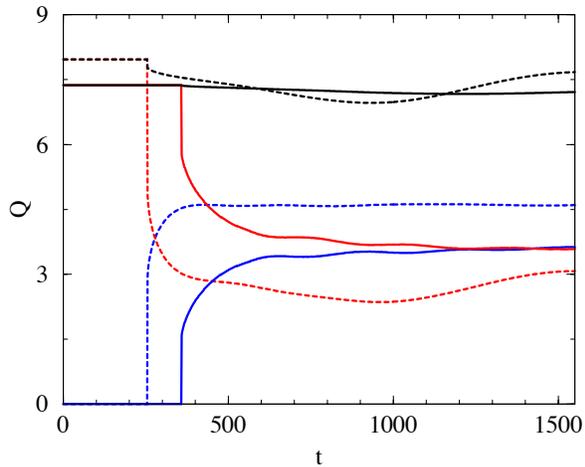}
\caption{Time evolution of the charge in the scalar field, $Q_{\rm SF}$ (red curves), in the BH (computed at the AH), $Q_{\rm BH}$ (blue curves), and of the total charge $(Q_{\rm T})$ (black curves), for model 8. Dashed curves correspond to scheme IIa (5\% perturbation) and solid lines to scheme IIb (1\% perturbation).}
\label{fig:graf8}
\end{minipage}
\end{figure}

The charge transfer from the scalar field (soliton) to the BH in scheme II is 
monitored in Fig.~\ref{fig:graf8}, where we plot the evolution of the charge of 
the BH, of the scalar field, and of the total charge, for model 8 and for the 
two perturbations considered (schemes IIa and IIb). A clear charge transfer can 
be seen from the scalar field to the BH at the time of formation of the AH. The 
total charge, however, decreases when the BH forms (black curves), signaling an 
issue. This is in fact due to the amount of perturbation we set up initially. 
Comparison of both schemes shows that when the initial perturbation to the 
solution is reduced from 5\% (scheme IIa, dashed black lines) to 1\% (scheme 
IIb, solid black lines), then total charge conservation holds at a higher 
accuracy. Using a large perturbation has the desirable feature that it triggers 
faster dynamics and hence leads to shorter simulations; however, as this example 
shows, too large perturbations introduce also undesirable features in the 
dynamics of the system, as total charge variation. 

\begin{figure}[t]
\begin{minipage}{1\linewidth}
\vspace{-0cm}\includegraphics[width=1.03\textwidth]{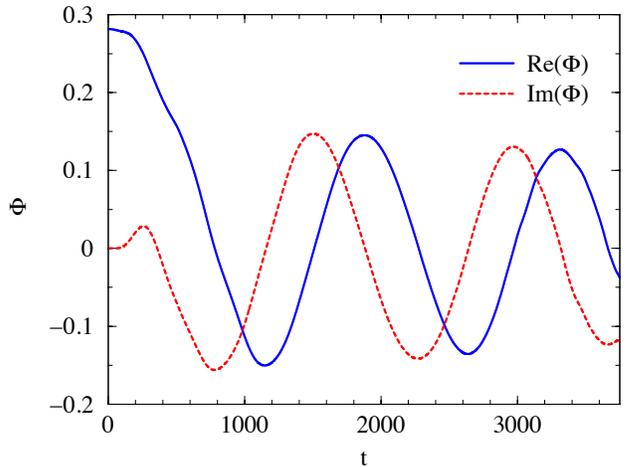}
\caption{Time evolution of the real (blue solid line) and imaginary (red dashed line) parts of the scalar field amplitude for model 8 and scheme IIb. Data are extracted at $r=5$.}
\label{fig:graf9}
\end{minipage}
\end{figure}

\begin{figure}[t]
\begin{minipage}{1\linewidth}
  \vspace{-0cm}\includegraphics[width=1.03\textwidth]{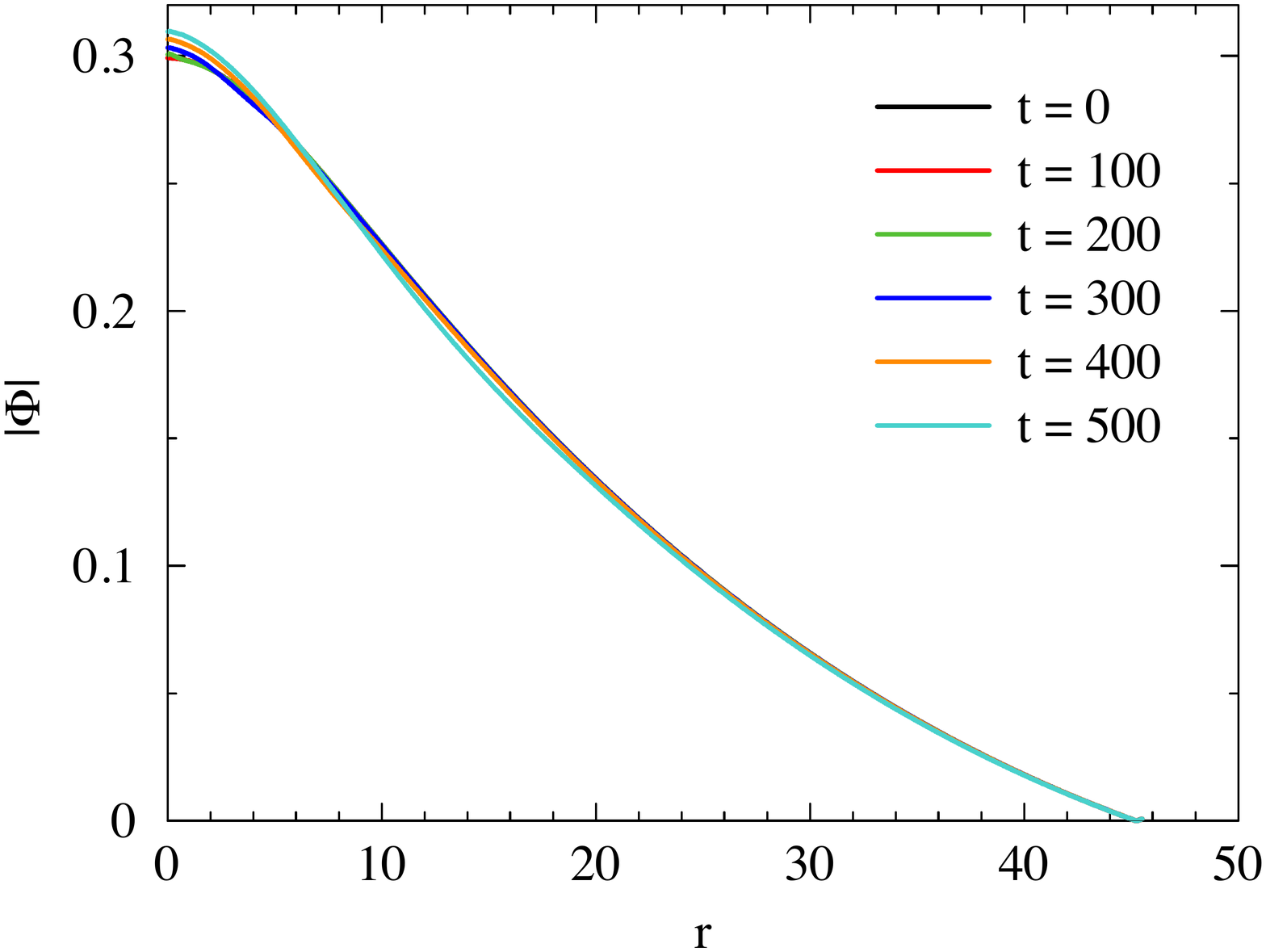}
  \hspace{-0.5cm}\includegraphics[width=1.03\textwidth]{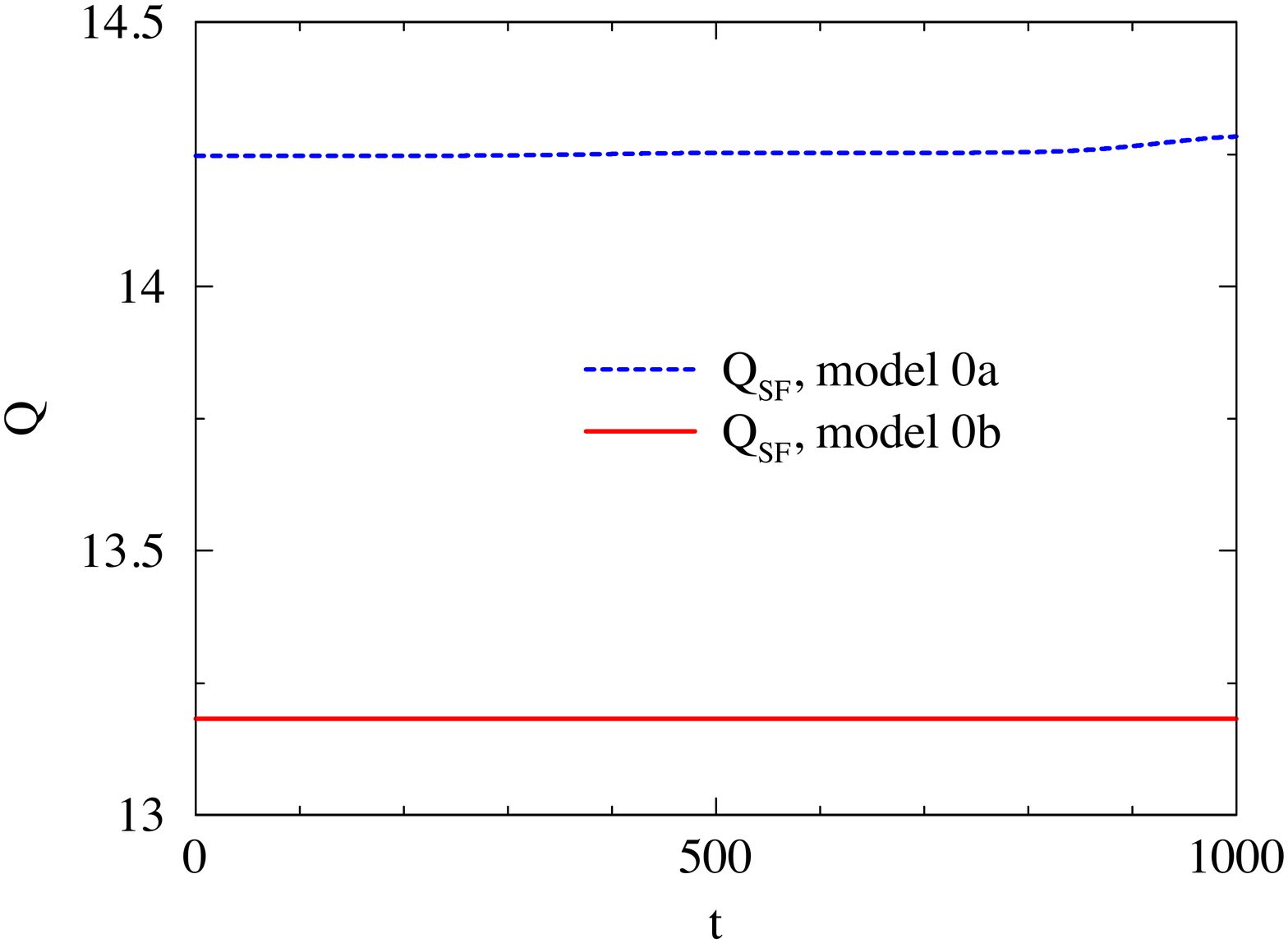}
\caption{(Top panel): Radial profile of the scalar field magnitude for model 0 
without adding a perturbation and for different times. (Bottom panel) : Time 
evolution of the scalar field charge for models 0a (blue dashed line) and 0b 
(red solid line). 
The total charge variation observed for model 0a towards the end of the 
simulation is related to the use of much too large a perturbations, as discussed 
in the main text.}
\label{fig:graf11}
\end{minipage}
\end{figure}

In Fig.~\ref{fig:graf9}, we show the scalar field amplitude extracted at $r=5$ for model 8 for scheme IIb. As before we find that  the real and imaginary parts of the scalar field oscillate with opposite phases. Moreover, as shown in Table~\ref{table1}, the oscillation frequency of the scalar field matches, to a good accuracy, the critical frequency, establishing the formation of a hairy BH at the threshold of superradiance. We observe that, for scheme IIb, the oscillation frequency is an order of magnitude smaller than in scheme I, so we can only capture roughly one period. We note that this already required using a larger computational grid than for the simulations based on scheme I. 

Let us focus now in the stable solution, model 0. We have already shown in Fig.~\ref{fig:graf7} that the evolution of the scalar field energy in this model is stable in scheme II, contrary to what we found when using scheme I. To add further proof of its
stable character we plot in Fig.~\ref{fig:graf11} the radial profile of the scalar field magnitude for different times for model 0 without initial perturbation (top panel) and the evolution of the scalar field charge (bottom panel). These results show that this model is indeed stable during our numerical evolutions with scheme II. The solution remains essentially invariant, up to numerical error and the constraint violations induced by the presence of the mirror.

\begin{figure}
\begin{minipage}{1\linewidth}\hspace{1.5cm}\includegraphics[width=1.03\textwidth, height=0.3\textheight]{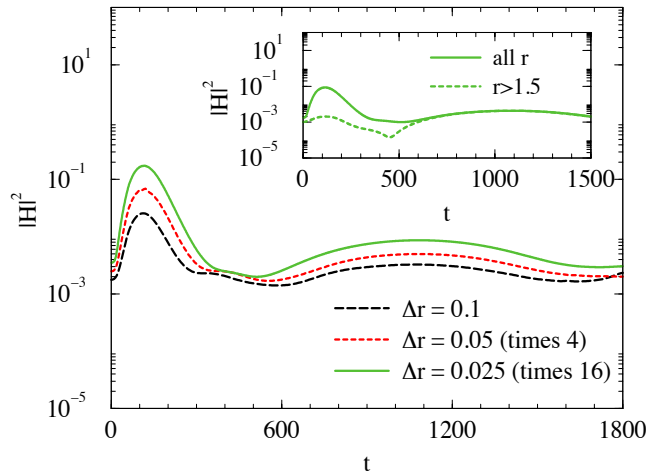}
\caption{Time evolution of the L2 norm of the Hamiltonian constraint for model 8 using scheme IIa.
The inset shows the evolution for the entire radial domain (solid line) and for $r>1.5$ (dahsed line), 
thus removing the radial zones that eventually lie within the AH.
}
\label{fig:graf12}
\end{minipage}
\end{figure}

Finally, in Fig.~\ref{fig:graf12} we show the evolution of the L2 norm of the 
Hamiltonian constraint for scheme IIa for model 8, for three radial grid 
spacings. We observe that although there is an initial growth of the Hamiltonian 
constraint violation, similarly to what happens with scheme I and also around 
the time the black hole forms, this violation now reaches a lower maximum than 
that observed in scheme I. Moreover, as in scheme I, 
the violations are significantly dissipated for $t\gtrsim 400$ and the evolution 
drives the values of the constraints to acceptable levels. Finally, the 
analysis with various grid spacings shows that the convergence order of the code 
is second order to reasonable accuracy, as expected. 

\section{Discussion}
\label{section4}

The endpoint of the superradiant instability of the Kerr BH, 
triggered by a confined bosonic scalar field, remains as an important open 
question in the theoretical understanding of BHs -- see~\cite{Brito:2015oca} for 
a review of superradiance. In particular, the discovery of new stationary BH 
solutions -- Kerr BHs with scalar~\cite{Herdeiro:2014goa,Herdeiro:2015gia} or Proca 
hair~\cite{Herdeiro:2016tmi} -- that exist at the threshold 
of superradiance, raises the question if these could be endpoints of the 
instability. 

As a simpler model for this problem, with important technical advantages, 
recent studies of the fully non-linear evolution of the superradiant 
instability have considered the EMcKG system and focused on the case of a 
RN BH enclosed in a cavity~\cite{Sanchis-Gual:2015lje,Sanchis-Gual:2016tcm}, 
which is also superradiantly unstable in the presence of a charged scalar 
field~\cite{Herdeiro:2013pia,Hod:2013fvl,Degollado:2013bha,Hod:2015hga,
Hod:2016kpm}.\footnote{See~\cite{Bosch:2016vcp} for the non-linear evolution of a superradiantly unstable RN BH in asymptotically Anti-de-Sitter spacetime.} The dynamical evolution of this unstable system has been shown to lead to the formation of hairy BHs at the threshold of superradiance. These BHs were found in~\cite{Dolan:2015dha} (see also~\cite{Basu:2016srp}), and can be faced as the analogues, in this system, of Kerr BHs with scalar hair~\cite{Herdeiro:2014goa,Herdeiro:2015gia}.

The existence of charged scalar solitons in this same system, recently reported in~\cite{Ponglertsakul:2016wae,Ponglertsakul:2016anb}, some of which are unstable, raises the question about the endpoint of these unstable solitons. Understanding this question has been the main motivation of this paper. We have tackled the question by considering two different dynamical schemes, that impose the confinement rule for the scalar field in the cavity differently. Our main conclusion is that, albeit there are quantitative differences between these two schemes -- most notably,  the end state of some of our models is a hairy BH when using scheme I,  and a bald RN BH when using scheme II --, the main message is similar. Indeed, we observe the same pattern for both schemes: there seems to be a critical value of $a_{0}$ for a given $\phi_{0}$ below which the collapse of unstable solitons leads to a hairy BH rather than to a bald RN BH. Models 6-10 in particular provide a good example. The value of the critical $a_{0}$ differs, but we see the two possibles outcomes for both schemes; for scheme I the transition occurs from model 9 to 10, whereas for scheme II it occurs from model 8 to 9. In a nutshell, we believe our numerical results provide evidence that there is a second channel for the dynamical formation of a hairy BH at the threshold of superradiance: from the decay of certain unstable solitons. The first channel is, of course, the aforementioned growth of the superradiant instability.

\bigskip

There are a number of interesting physical questions in this system, which our study raises, for instance:
\begin{description}
\item[1)] is there a simple criterion/physical understanding separating the decay into a hairy BH {\it vs} a bald BH? We have observed that for a fixed value of $\phi_0$ sufficiently small (large) values of $a_0$ produce a hairy (bald) BH. Can one get a deeper understanding?
\item[2)] There are some stable hairy BHs that cannot form through the 
first channel, as they do not obey the $\mu<q$ superradiance 
requirement~\cite{Herdeiro:2013pia} or even a stronger lower 
bound~\cite{Hod:2016kpm} -- see the discussion in~\cite{Ponglertsakul:2016anb}. 
Can these form from 
the evolution of an unstable soliton? More generically, what is the impact of the scalar field mass $\mu$ on our results?
\item[3)] The hairy BHs that form through the ``old" channel 
(superradiant growth around a bald BH) are not very hairy, in the sense 
that the mass of the scalar configuration is not comparable to the mass of the 
BH. Could very hairy BHs, with almost all of the energy in the hair, form from 
the decay of unstable solitons?
\end{description}

Whereas all these questions are certainly worthwhile to address, our study has 
also exposed some frailties of this system that raise doubts on how much more 
one can pursue it, without introducing further complexity. The 
scalar-field-confining mirror did not impose important challenges when 
addressing \textit{test} scalar field 
evolutions~\cite{Herdeiro:2013pia,Hod:2013fvl,Degollado:2013bha,Hod:2015hga,
Hod:2016kpm}\footnote{For test fields, in fact, the confining mirror has been 
widely used in other studies of fields interacting with charged BHs, see 
$e.g.$~\cite{Li:2014xxa,Li:2014gfg,Li:2014fna,Li:2015bfa,Sakalli:2015uka,
Li:2015mqa,Huang:2016zoz,Hod:2016nhu}.}, or fully non-linear \textit{stationary} 
solutions~\cite{Dolan:2015dha,Ponglertsakul:2016wae,Ponglertsakul:2016anb,
Basu:2016srp}. When making fully non-linear dynamical evolutions, however, the 
``virtual" mirror becames a source of constraint violations. In our previous 
simulations~\cite{Sanchis-Gual:2015lje,Sanchis-Gual:2016tcm}, the scalar field 
was initialized with very small values and never became too large. In the case 
under study here, however, the scalar field is large from the outset, as it 
describes a self-gravitating scalar soliton, which makes the mirror a more 
important source of  numerical error and, consequently, makes the evolutions 
more challenging. Moreover, in the horizonless case, the coordinate singularity 
(of spherical coordinates) at the origin also sources constraint violations. 
Fortunately, we have observed that the system evolves towards acceptable values 
of the constraint violations. 

\bigskip

We close by returning to the parallelism with the asymptotically flat case. The ``cousin" solutions to the solitons studied herein are the well known \textit{boson stars}~\cite{Schunck:2003kk}, self-gravitating systems formed by massive scalar fields (eventually, but not necessarily, self-interacting).  There are stable and unstable boson stars. Stable ones undergo
oscillations when perturbed and show similar properties to other 
compact objects, such as neutron stars.  
Unstable boson stars, however, may collapse into BHs, radiate the energy excess or completely disperse away. These behaviours have been explicitly shown by numerical evolutions for spherically symmetric boson stars~\cite{Seidel:1990jh,Alcubierre:2001ea,Guzman:2004jw,Bernal:2006it,Bernal:2009zy}. In that case, however, there are no hairy BHs~\cite{Pena:1997cy}, which require rotation and thus exist only in axisymmetry~\cite{Herdeiro:2014ima}. Thus, in a similar spirit to the results in this work, could unstable rotating boson (or Proca) stars form Kerr BHs with scalar (or Proca) hair? This is an interesting open question.

\section*{Acknowledgements}

This work has been supported by the Spanish MINECO (AYA2013-40979-P), 
by the Generalitat Valenciana (PROMETEOII-2014-069), by the 
CONACyT-SNI-M\'exico, by the FCT (Portugal) IF programme, by the CIDMA (FCT) 
strategic project UID/MAT/04106/2013 and by the EU grants  
NRHEP--295189-FP7-PEOPLE-2011-IRSES and H2020-MSCA-RISE-2015 Grant No. 
StronGrHEP-690904. Computations have been 
performed at the Servei d'Inform\`atica de la Universitat de Val\`encia and at the Blafis cluster at the University of Aveiro.


\newpage

\bibliography{num-rel}

\end{document}